\documentclass[lnicst]{svmultln}
\usepackage{amssymb}
\usepackage{graphicx}

\usepackage{url}

\usepackage[pdfpagelabels,hypertexnames=false,breaklinks=true,bookmarksopen=true,bookmarksopenlevel=2]{hyperref}

\begin{document}

\mainmatter  % start of an individual contribution

\title{Fingerprint for Network Topologies}

\titlerunning{Fingerprint for Network Topologies}

\author{Yuchun Guo \inst{1} \and Changjia Chen \inst{1}
\and Shi Zhou\inst{2} }
\authorrunning{Y.~Guo and C.~Chen and S. Zhou}

% the affiliations are given next
\institute{Beijing Jiaotong University, China\\
\email{ychguo@bjtu.edu.cn, changjiachen@sina.com}\\
\and
University College London, United Kingdom\\
\email{s.zhou@adastral.ucl.ac.uk}\\}

\toctitle{Fingerprint for Network Topologies}

\tocauthor{Authors' Instructions} \maketitle

\begin{abstract}

A network's topology information can be given as an adjacency matrix. The {bitmap of sorted adjacency matrix}
(BOSAM) is a network visualisation tool which can emphasise different network structures by just
\emph{looking} at reordered adjacent matrixes. A BOSAM picture resembles the shape of a flower and is
characterised by a series of `leaves'. Here we show and mathematically prove that for most networks, there is
a self-similar relation between the envelope of the BOSAM leaves. This self-similar property allows us to use
a single envelope to predict all other envelopes and therefore reconstruct the outline of a network's BOSAM
picture. We analogise the BOSAM envelope to human's fingerprint as they share a number of common features,
e.g.~both are simple, easy to obtain, and strongly characteristic encoding essential information for
identification.


\end{abstract}

\section{Introduction}

During the last decade there has been an international effort to
understand the structure and dynamics of complex networks in social,
biological, and technology systems~\cite{wasserman94, watts99,
Barabasi99, strogatz01, albert02, Maslov02a, bornholdt02, newman03a,
dorogovtsev03a, Pastor04, boccalettia06, newman06}. These networks
are  very large, containing thousands or even millions of entities
(nodes) interacting with each other (links), and their structures
are irregular, evolving and inherently stochastic. The statistical
physics methods have been widely used in studying complex networks.

Complementary to this effort, a number of network visualisation tools have been
proposed to illustrate network topologies, such as~\cite{alvarez05, Guo07a,
Chakrabarti07}. These techniques take the advantage of human being's extraordinary
ability in recognising patterns in images and therefore allow us to compare two
networks by \emph{seeing} whether the networks visualisations \emph{look} similar to
each other.

Of our particular interest is a tool called the \emph{bitmap of sorted adjacency
matrix} (BOSAM)~\cite{Guo07a}. It sorts a network's nodes in a specific order such
that the bitmap representation of the reordered adjacency matrix resembles a
`flower'. The shape of the flower reveals many topological properties of the
network. A BOSAM flower consists of a series of `leaves', each of which is
characterised by its envelope.

In this paper we demonstrate and mathematically prove that the way
the adjacency matrix is reordered for BOSAM gives rise to a
self-similar relation between the envelopes of the leaves. This
self-similar property allows us to use just one envelope to predict
all other envelopes and therefore recover the shape of a BOSAM
flower. We also show that if a network preserves its macroscopic
structure during the network growth, the BOSAM envelopes scale with
the network's size. We remark that an envelope of a network's BOSAM
is analogous to a fingerprint of a human being, which is a small
token, easy to obtain, valid for life, and encodes essential
information for identification.

\section{Bitmap Of Sorted Adjacency Matrix}\label{section:rules}

For a network with $N$ nodes, the connectivity information between the nodes can be
given as an $N*N$ adjacency matrix, in which entry $a_{ij}$ is the number of links
connecting between nodes with indexes $i, j \in\{1, 2, . . . , N\}$. For an
undirected, simple network (no self-loop or repeat link), the adjacency matrix
becomes a symmetric $(0, 1)$-matrix with zeros on its diagonal, where entry $a_{ij}$
is mirrored by entry $a_{ji}$. This matrix can be represented as a black-and-white
bitmap, i.e.~if $a_{ij}=1$, a black pixel is placed at the coordinate of $(i, j)$;
otherwise a white pixel is placed there. One can see such a bitmap is not very
helpful if node indexes are randomly assigned.

Degree $k$ is defined as the number of links a node has. For a given
network, we sort nodes in ascending order of the degree. For nodes
having the same degree, we arrange them in ascending order of the
largest neighbor degree, $\omega$, which is the largest degree of a
node's neighbours. For nodes having both the same degree and the
same largest neighbor degree, we reorder them in ascending order of
the largest neighbor index, $e$, which is the largest index of a
node's neighbours. We then reassign each node a new index using the
node's position in the sorted list. Then, for two nodes with indexes
$i<j$, we have one of the followings:
\begin{itemize}
\item $k_i< k_{j}$; \item $k_i = k_{j}$ and $\omega_i<\omega_{j}$; \item $k_i = k_{j}$,
$\omega_i=\omega_{j}$ and $e_i<e_{j}$.
\end{itemize}

The above node sorting rule produces a reordered adjacency matrix,
whose bitmap visualisation is called the \emph{bitmap of sorted
adjacency matrix} (BOSAM). Fig.\,\ref{fig:BOSAM} shows BOSAMs for
six networks. The names and sources of the datasets for these
networks are given in Table 1. Each BOSAM picture resembles a
`flower' which consists of a series of `leaves' symmetrically
arranged along the bitmap's diagonal. The shape of the flower
reflects a number of network topological properties. For simplicity,
in the following we only discuss vertical leaves above the diagonal.

\begin{table*}
\begin{center}
\caption{Properties of networks under study: (\textbf{a})~the
Erd\"{o}s-R\'enyi (ER) model~\cite{erdos59} which generates random
networks having a Poisson degree distribution; (\textbf{b})~the
Barab\'asi-Albert (BA) model~\cite{Barabasi99} which generates
scale-free networks with a power-law degree distribution;
(\textbf{c})~the Positive-Feedback Preference (PFP)
model~\cite{zhou04d, zhou06b,zhou07a} which generates Internet-like
networks; (\textbf{d}) the scientific collaboration
network~\cite{newman01a,newman01b}, in which nodes represent
scientists and a connection exists if they coauthored at least one
paper in the e-print archive {http://xxx.lanl.gov/archive/cond-mat/}
from 1995 to 1998; (\textbf{e})~the protein interaction
network~\cite{Maslov02a,colizza05}, in which nodes represent
proteins in the yeast \emph{Saccharomyces cerevisiae}
({http://dip.doe-mbi.ucla.edu/}) and a connection exists if they
interact with each other; (\textbf{f})~the Internet network at the
autonomous system (AS) level~\cite{Pastor04, mahadevan05b} based on
the traceroute data collected by CAIDA in April 2003~\cite{itdk0403,
CAIDA}; and (\textbf{g, h})~the Internet AS networks based on the
BGP data collected by the Oregon Route Views project
(http://www.routeviews.org/) in  October 2001 and September 2006,
respectively. On the AS Internet, nodes represent Internet service
providers and a connection exists if they have a commercial
agreement to exchange traffic. The shown properties are: the number
of nodes $N$ and links $L$, the average node degree $\langle
k\rangle=2L/N$, and the characteristic node degree $k^*$ which has
the largest degree distribution.} \label{table:networks}
\renewcommand{\tabcolsep}{1pc} % enlarge column spacing
\renewcommand{\arraystretch}{1.5} % enlarge line spacing
\begin{tabular}{lrccc}
& & & \\
\hline\hline
 & $N$  &  $L$      &  $\langle k\rangle$   & $k^*$\\
\hline
(\textbf{a})~ER  network &10,000& 30,000 & 6 & 6 \\
(\textbf{b})~BA  network &10,000& 30,000 & 6 & 3 \\
(\textbf{c})~PFP  network &10,000& 30,000 & 6 & 2 \\
(\textbf{d})~Scientific collaboration  &15,179& 43,011 &  5.7& 2\\
(\textbf{e})~Protein interaction &4,713& 14,846 & 6.3 &1 \\
(\textbf{f})~Internet (traceroute) &9,204& 28,959 & 6.3 & 2  \\
(\textbf{g})~Internet (BGP-2001) &12,033& 21,742 & 3.6 & 1 \\
(\textbf{h})~Internet (BGP-2006) &23,480& 49,077 & 4.2 &1\\
\hline\hline
\end{tabular}
\end{center}
\end{table*}

\subsection{Degree distribution}

The degree distribution $P(k)$ is the probability of finding a $k$-degree node in a
network. For a network with $N$ nodes, the number of $k$-degree nodes is
$N_{k}=N\cdot P(k)$, and the number of nodes with degrees equal to or smaller
than~$k$ as $N_{\leqslant k}=N\cdot \sum_1^k P(k)$. According to the node sorting
rule of BOSAM, the indexes of $k$-degree nodes are $I_k = \{i\,|\,
N_{\leqslant{k-1}}<i\leqslant N_{\leqslant k}\}$.

Each leaf is associated with a node degree $k$ because it is formed by pixels
representing  connections linking to $k$-degree nodes. In other words, the leaf for
degree~$k$ represents entries $\{a_{i, j}=1\,|\,i\in I_k\}$ in the reordered
adjacency matrix. Thus the width of the $k$-degree leaf is $N_k$.

\begin{figure*}
\includegraphics[width=12cm]{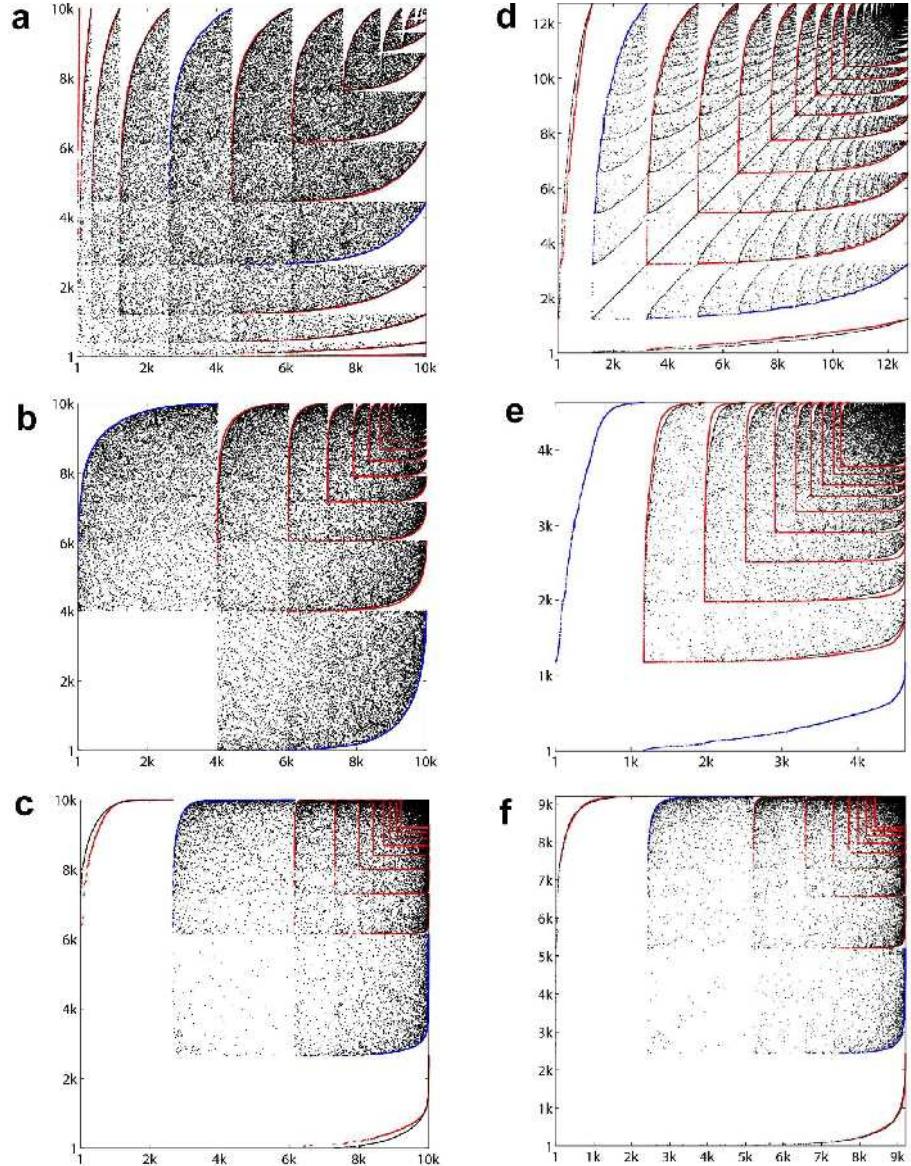}
\caption{Bitmap of sorted adjacency matrix (BOSAM) for (\textbf{a})~the ER model, (\textbf{b})~the BA model,
(\textbf{c})~the PFP model, (\textbf{d})~the scientific collaborations, (\textbf{e})~the protein
interactions, and (f)~the AS Internet. For each network, the envelope which is used as the root for the
prediction is shown in blue colour and the predicted envelopes are shown in red colour.}\label{fig:BOSAM}
\end{figure*}

In Fig.\,\ref{fig:BOSAM},  the widest leaf in the ER network's BOSAM is for degree
6, which reflects that the network has a Poisson degree distribution which peaks at
the average node degree of 6. Other networks are `scale-free' having a power-law
degree distribution~\cite{Barabasi99}. This means most nodes are low-degree nodes,
whereas a small number of nodes have very large degrees. This is reflected on BOSAMs
as the width of leaves decreases rapidly with the node degree.

\subsection{Degree-degree correlation \label{section:knn}}

Degree-degree correlation is a widely studied property~\cite{vazquez03,
Mahadevan06}.  The protein interaction network, the Internet and the PFP network
have a negative degree-degree correlation, or so-called disassortative
mixing~\cite{newman03}, which means low-degree nodes tend to connect with
high-degree nodes and vice versa. This is reflected on their BOSAMs in
Fig.\,\ref{fig:BOSAM} as pixels are densely distributed along the upper envelope of
the leaves. In contrast, the scientific collaboration network exhibits a positive
degree-degree correlation, or assortative mixing~\cite{newman02}, which means nodes
tend to connect with alike nodes of similar degrees. This is characterised on the
BOSAM as a series of lines are radiated from the top-right corner across the leaves.
The ER network and the BA network have a neutral degree-degree correlation, which is
illustrated on the BOSAMs as pixels are fairly evenly distributed on the leaves.

\subsection{Rich-club}

In the Internet and the PFP model, the high-degree nodes, `rich' nodes, are tightly
interconnected with themselves, forming a rich-club~\cite{Zhou04a, zhou07b}. This is
reflected on their BOSAMs as the top-right corner is almost fully covered by pixels.
This is not the case for the ER network and the BA network where high-degree nodes
are sparsely interconnected with themselves.

\vspace{5mm}

In summary, BOSAM provides a simple and effective way to emphasising different
network structures. We can compare network topologies by just \emph{looking} at
their BOSAMs. For example one can see that although the BA model has been widely
used as a generic model for all scale-free networks, the model does not closely
resemble the Internet, the protein interaction and the scientific collaborations. In
fact the three real networks themselves are different from each other in profound
ways. The PFP model well resembles the Internet based on the traceroute data (see
Table 1).

\section{Self-similar property of BOSAM}

One improvement to the previous version of BOSAM~\cite{Guo07a} is
that here we consider the largest neighbor index $e$ as well in the
node sorting rule. This reduces zigzag in BOSAM and as a result the
envelopes of the leaves become smooth, solid curves.

The envelope of the leaf for degree $k$ consists of $N_k$ pixels given as
\begin{equation}
E_k=\{(i, e_i)|i\in I_k\},
\end{equation}
where $e_i$ is the largest neighbor index of node $i$ and $I_k$ is
the set of indexes of $k$-degree nodes. According to the node
sorting rule of BOSAM (see Section~\ref{section:rules}), the degree
of node $e_i$ is the largest neighbor degree of node $i$,
i.e.~$k_{e_i}=\omega_i$. One can see that the envelope $E_k$ is
given by the cumulative distribution function $F_k(\omega)$, which
is the probability for a $k$-degree node having the neighbours
largest degree less than or equal to $\omega$.

\subsection{Self-similar relation between BOSAM envelopes}

\textbf{Theorem 1.} \emph{In a network, if a node's neighbours degree  $\mu$ is independent and identically
distributed (\emph{i.i.d.}), the cumulative distribution functions $F_k(\omega)$ and $F_l(\omega)$ for
$k$-degree nodes and $l$-degree nodes, respectively, have the following self-similar relation,}
\begin{equation} F_k(\omega) = F_l(\omega)^{^k/_l}.  \label{eq: self-similar}
\end{equation}

The proof of Theorem~1 is given in Appendix I. This self-similar property of BOSAM allows us to use just one
envelope, we call it the root envelope, to predict all other envelopes (see the self-similar algorithm in
Appendix III).

The characteristic degree $k^*$ is the node degree having the largest number of
nodes, i.e.~$N_{k^*}\geqslant N_k$ or $P(k^*)\geqslant P(k)$. The envelope of the
leaf for the characteristic degree contains more information than other envelopes.
Fig.\,\ref{fig:BOSAM} illustrates the prediction result. For each network, we use
the envelope for the characteristic degree (see Table~1) as the root envelope. We
highlight the root envelope in blue colour and the predicted envelopes for other
degrees in red colour. One can see that the predicted envelopes well overlap with
the real envelopes beneath them.

\subsection{Discussion}

The self-similar relation between BOSAM envelopes is different from other scaling  properties in networks,
such as the scaling property of community size in social networks~\cite{Guimera03}. As shown in the proof of
Theorem 1, the self-similar relation between BOSAM envelopes is originated from the way we reorder the
adjacency matrix, and therefore it is valid for all networks, regardless of networks degree distribution or
degree-degree correlation.

The only condition for the proof of Theorem 1 is that the neighbours degree $\mu$ is independent and
identically distributed (\emph{i.i.d.}). One should not confuse this condition with the degree-degree
correlation property of a network. The \emph{i.i.d.} condition means that a network's degree-degree
correlation (whether the correlation is negative or positive) is consistent for all nodes. We can infer
whether a network is \emph{i.i.d.} by observing whether the prediction of BOSAM envelopes is accurate. Fig.~1
shows that most networks under study satisfy the \emph{i.i.d.} condition. By comparison, the predicted
envelopes for the protein interaction network do not precisely (but still quite closely) match the real
envelopes. This suggest that the protein network are not strictly \emph{i.i.d.}.

\subsection{Scaling of BOSAM envelopes}

\begin{figure*}
\includegraphics[width=12cm]{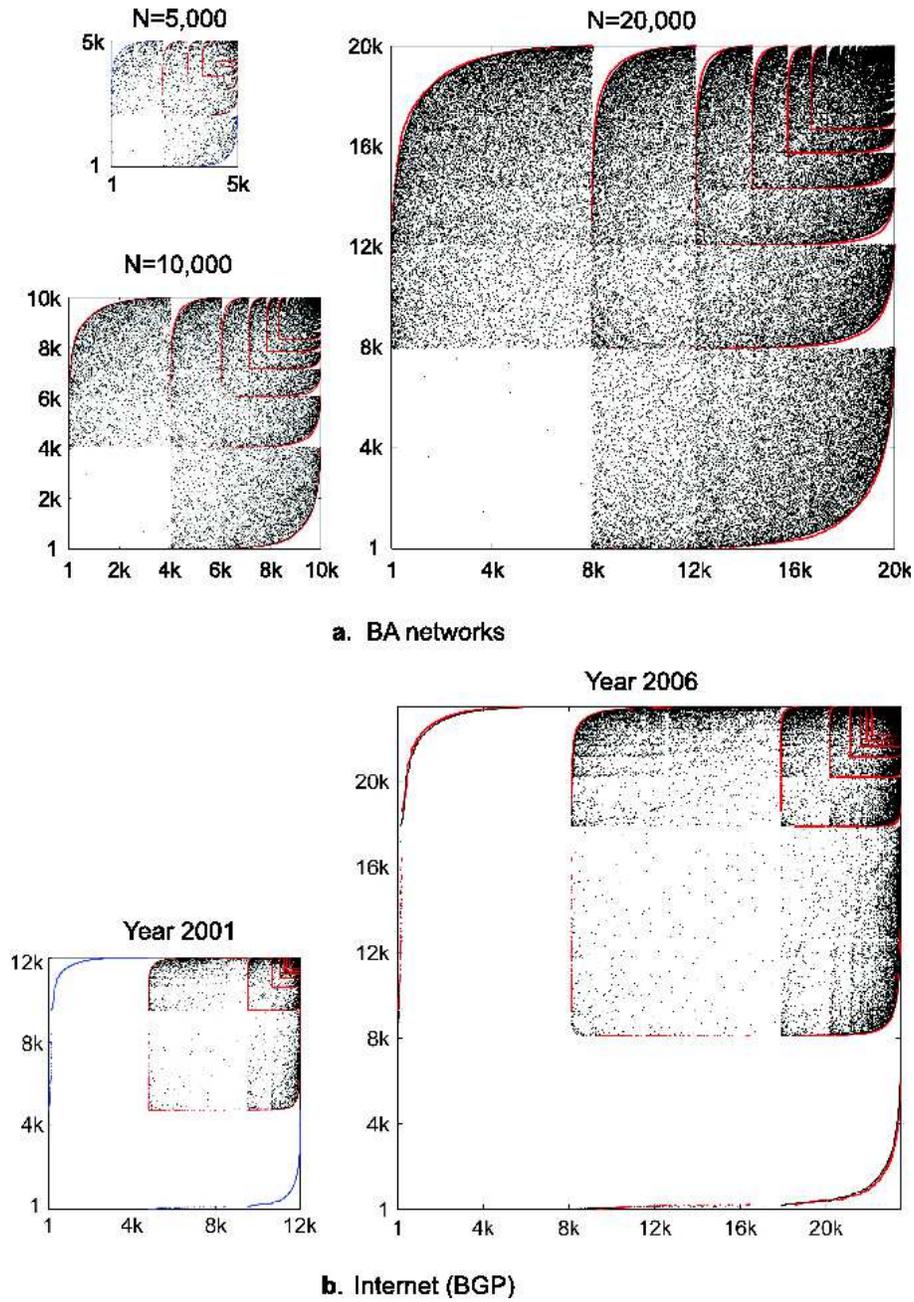}
\caption{Prediction of BOSAM envelopes for growing networks.
\textbf{(a)} BA networks with the number of nodes $N=5,000$,
$N=10,000$ and $N=20,000$. \textbf{(b)} Internet networks based on
BGP data collected in year 2001 and 2006 (see Table~1). In
\textbf{(a)} and \textbf{(b)}, red envelopes in larger networks are
predicted from the blue envelope in the smallest network. }
\label{fig:SizeScaling}
\end{figure*}

If two networks have the same macroscopic structure, their BOSAM pictures should
look the same. This should be the case even if the networks are of different sizes.
For example it is known that the BA model preserves its  topological structure
during network growth~\cite{Barabasi99}. Therefore, as shown in
Fig.\,\ref{fig:SizeScaling}-(a), the BOSAM pictures for the three BA networks with
different sizes indeed look the same. We can use the scaling algorithm in Appendix
IV to accurately predict the envelopes  in the two larger networks (in red colour)
from the envelopes in the small network (which themselves are predicted from the
root envelope in blue colour).  Thus the scaling property of BOSAM envelopes can be
used to test whether two networks with different sizes have the same macroscopic
structure.

Fig.\,\ref{fig:SizeScaling}-(b) shows the BOSAM pictures for the Internet networks
based on the BGP data collected in 2001 and 2006 (see Table~1). The envelopes in the
large network are precisely predicted by scaling the envelopes in the small network.
This suggests that during the 5-year period, although the Internet doubled its size,
it well preserves its macroscopic structure.

\section{BOSAM envelope as network fingerprint}

Based on the above observations, we remark the analog between BOSAM envelope and
human being's fingerprint in the following ways. (1)~ A BOSAM envelope is a small
token of the network's adjacency matrix, in the form of a set of coordinates
$E_k=\{(i, e_i)|i\in I_k\}$. Such a relatively small amount of information is easy
to obtain, store and process. (2)~A single BOSAM envelope is able to recover all
other envelopes and thus provide an outline description of a network's BOSAM.
(3)~The envelope fingerprint is valid for a growing network as far as the network
preserves its macroscopic structure, just like a person's fingerprint is valid for
life. And (4)~A BOSAM envelope contains essential information that characterises the
network's topology.

\begin{figure*}
\includegraphics[width=10cm]{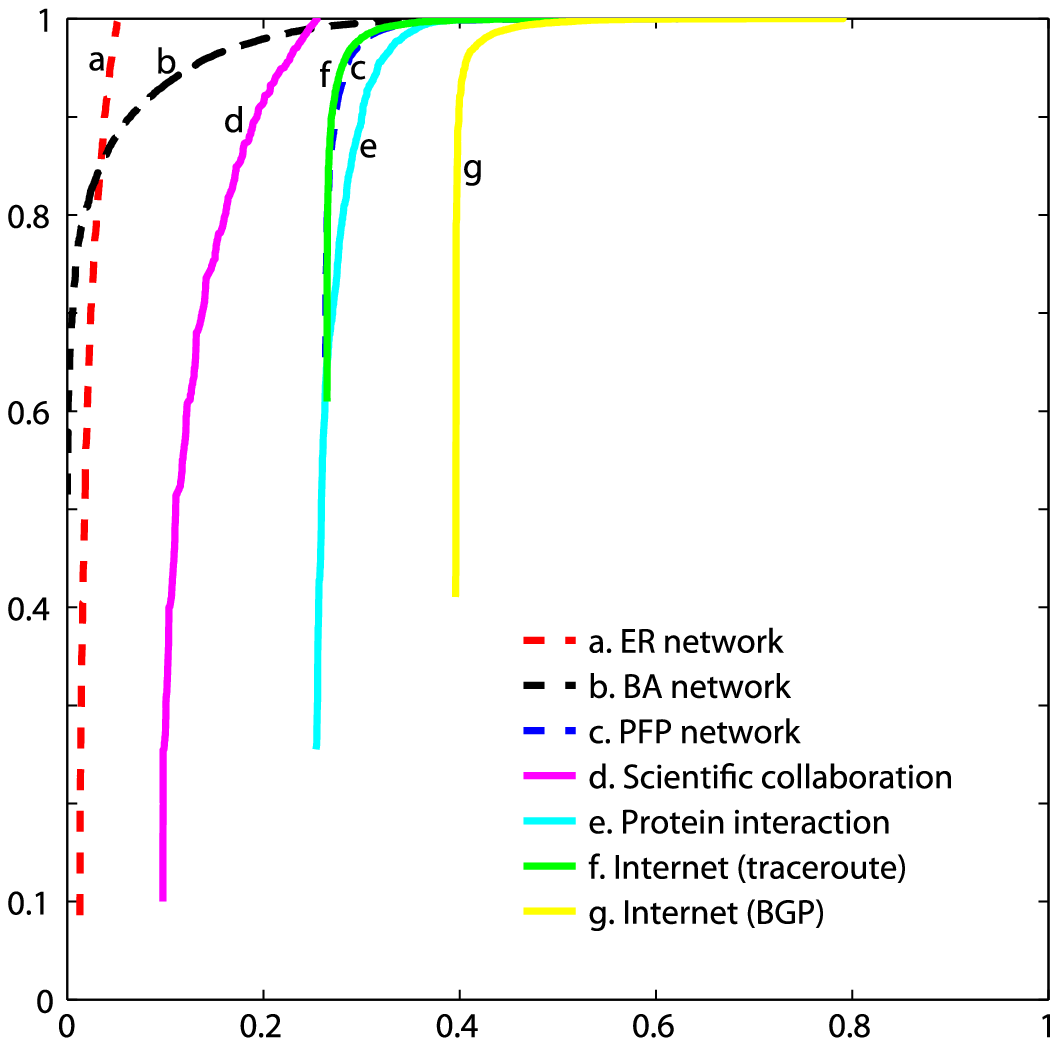}
\caption{BOSAM fingerprint for all networks under study. Each
envelope is normalised by the number of nodes in its own network.}
\label{fig:FingerPrint}
\end{figure*}

Fig.\,\ref{fig:FingerPrint} shows one envelope fingerprint for each of the networks
under study. For comparison purpose, all the envelopes shown are of the leaves for
the node degree 2, except for the BA network which does not contain 2-degree nodes
and therefore the envelope for degree 3 is shown instead. The size of the envelopes
are normalised by the number of nodes in the networks. We can see the envelope
fingerprint of the networks are strongly characteristic.

Fig.\,\ref{fig:FingerPrint} shows two fingerprints for Internet networks based on
different data sources~\cite{mahadevan05b}. The fingerprint for the traceroute
Internet (line 6) is positioned to the left of that for the BGP Internet (line 7).
This reflects one of the key differences between the two data sources that 1-degree
nodes count for a larger proportion in the BGP data than in the traceroute data.
However the two fingerprints have the same shape, which suggests the fact that the
macroscopic structure of the two Internet networks are similar. The close
approximation between the traceroute Internet and the PFP network (line 3) is
evidently shown by the close match of their fingerprints. One would expect that
minor revision would enable the PFP model to resemble the BGP Internet as well.

\section{Conclusion}

BOSAM is a visualisation tool for network topologies. The simple tool provides an
effective way to emphasise networks topological differences or similarities by just
looking at the bitmap of reordered adjacency matrixes.

A network's BOSAM is characterised by a series of leaves and the shape of the leaves are described by their
envelopes. We show there is a self-similar relation between the envelopes for most networks. This properties
allow us to use one single envelope to reconstruct all envelopes. For an evolving network which preserves its
structure, the BOSAM envelopes scale with the growing size of the network. In these respects we suggest that
the BOSAM envelope can be used as a self-similar fingerprint for network topologies.

\subsubsection*{Acknowledgments.} Y.~Guo and C.~Chen are supported by the Natural Science Foundation of China
under grant no.~60672069 and 60772043, and the National Basic Research Program
(973 Program) of China under grant no.~2007CB307101. S.~Zhou is supported by
the Royal Academy of Engineering and the UK Engineering and Physical Sciences
Research Council (EPSRC) under grant no.~10216/70.

\section*{APPENDIX I ~~Proof of Theorem 1}

For a $k$-degree node, the neighbours degree $\mu$ consists of $k$
variants $\mu_1,\, \mu_2,\, ...,\, \mu_{k}\,$. We reorder them so
that $\nu_1\leqslant\nu_2\leqslant...\leqslant\nu_{k}$. Thus the
largest neighbor degree $\omega=\nu_k$.

According to the order statistic theory (see Appendix~II\,), the
probability function of $\omega$ for $k$-degree nodes is given by
\begin{equation}
P_k(\omega) = P_k(\nu_{k}) = k\,[F_k(\mu)]^{(k-1)}\,P_k(\mu)
\end{equation}
where $P_k(\mu)$ is the  probability function of $\mu$ and
$F_k(\mu)$ is the cumulative distribution function of $\mu$, for
$k$-degree nodes.

In a network where $\mu$ is \emph{i.i.d.}, we have $P_k(\mu)=P(\mu)$
and $F_k(\mu)=F(\mu)$. Then
\begin{equation}
P_k(\omega) = k\,[F(\mu)]^{(k-1)}\,P(\mu).
\end{equation}

Therefore we have
\begin{eqnarray}
F_k(\omega) & = & \int_0^{\omega} P_k(\omega)\, d\omega\\
           & = & \int_0^\mu k\,[F(\mu)]^{(k-1)}\,P(\mu)\, d\mu\\
            & = & \int_0^\mu k\,[F(\mu)]^{(k-1)}\,{dF(\mu)\over d\mu}\, d\mu\\
            & = & \int_0^{F(\mu)} k\,[F(\mu)]^{(k-1)} \,d{F(\mu)}  \\
            & = & \int_0^{F(\mu)} \left( [F(\mu)]^k \right)' \,d{F(\mu)}  \\
            & = & [F(\mu)]^k.
\end{eqnarray}
Similarly we can have $F_l(\omega) = [F(\mu)]^l$ for $l$-degree
nodes. Thus Theorem 1 is proved:
\begin{equation}
F_k(\omega) = [F(\mu)]^k = [~[F(\mu)]^l~]^{^k/_l} =
[F_l(\omega)]^{^k/_l}.
\end{equation}

\section*{APPENDIX II ~~Order statistic}

Given a sample of $n$ variants $X_1, X_2..., X_N$, reorder them so
that $Y_1\leqslant Y_2\leqslant...\leqslant Y_N$. Then $Y_r$ is
called the $r^{\,th}$ order statistic~\cite{hogg70} for $r=1, 2,
..., N$.

If $X$ has the probability function $P(X)$ and the cumulative
distribution function (CDF) $F(X)$, then the probability function of
$Y_r$ is given by~\cite{rose02}
\begin{equation}
P(Y_r)={N!\over(r-1)!\,(N-r)!}\,\,[F(X)]^{r-1}\,[1-F(X)]^{N-r}\,P(X).\label{eq:OrderStatistic}
\end{equation}
Therefore the probability function of the largest value $Y_N$ is
\begin{equation}
P(Y_N)=N\,[F(X)]^{N-1}\,P(X).\label{eq:OrderStatistic-max}
\end{equation}

\section*{APPENDIX III ~~Algorithm for predicting envelopes for other degrees}

For a network with $N$ nodes and degree distribution $P(k)$, if we
know the BOSAM envelope $E_l=\{(i,\, e_i)\,|\,i\in I_l\}$ for degree
$l$, then we can use Theorem~1 to predict the envelope $E_k$ for
degree $k$ as
\begin{equation}
\tilde{E_k}=\{(\tilde{i}, e_i)\,|\,i\in {I}_l\} ,
\end{equation}
where
\begin{equation}
\tilde{i} = N_{ < {k}} + N_k \left( \frac {i-N_{ < {l}}} {N_l}
\right)^{^k/_l}
\end{equation}
and\\ \centerline{$N_{k}=N\cdot  P(k)$,~~ $N_{<k}=N\cdot
\sum_1^{k-1} P(k)$.}

\section*{APPENDIX IV ~~Algorithm for scaling envelopes with network size}

For a network with $N$ nodes, if we know the BOSAM envelope
$E_k=\{(i,\,e_i)\,|\,i\in I_k\}$ for degree $k$, the scaling of the
envelope to the network size of $N'$ is given by
\begin{equation}
\tilde{E'_k}=\{(\tilde{i'}, \tilde{e'_i})\,|\,i\in {I}_k\},
\end{equation}
where $$\tilde{i'}= i\,{N'\over N}$$ and $$\tilde{e'_i}
=e_i\,{N'\over N}.$$

%
%\pagebreak

\end{document}